\documentclass[a4paper,12pt]{article}
\usepackage[left=2cm,right=2cm,top=2cm,bottom=2cm]{geometry}
\usepackage{amsmath,booktabs,amsfonts,amssymb,latexsym,amsthm,verbatim,color,lscape}
\usepackage[figuresright]{rotating}
 
\usepackage{hyperref}
\hypersetup{colorlinks,
citecolor=black,
filecolor=black,
linkcolor=black,
urlcolor=black,
pdftex}
\usepackage{color}

\usepackage{graphicx}

\usepackage{natbib}
\begin{document}
\def\a{\alpha}
\newcommand{\bea}{\begin{eqnarray}}
\newcommand{\eea}{\end{eqnarray}}

\def\th{\theta}
\def\thbf{\boldsymbol{\theta}}
\def\dbf{\boldsymbol{\delta}}

\def\R{\mathbf{R}}
\def\x{\mathbf{x}}
\def\X{\mathbf{X}}
\def\Y{\mathbf{Y}}
\def\y{\mathbf{y}}
\def\a{\alpha}
\def\s{\mathbf{s}}
\def\R{\mathbf{R}}
\def\I{\mathbf{I}}
\def\x{\mathbf{x}}
\def\X{\mathbf{X}}
\def\bfv{\mathbf{v}}
\def\V{\mathbf{V}}
\def\Y{\mathbf{Y}}
\def\y{\mathbf{y}}
\def\bbf{\boldsymbol{\beta}}
\def\gbf{\boldsymbol{\gamma}}
\def\obf{\boldsymbol{\omega}}
\def\Obf{\boldsymbol{\Omega}}
\def\U{\mathbf{U}}
\def\J{\mathbf{J}}
\def\ga{\a}
\def\u{\mathbf{u}}
\def\C{\mathbf{C}}
\def\p{\mathbf{p}}
\def\1{\mathbf{1}}
\def\t{\mathbf{t}}
\def\W{\mathbf{W}}
\def\Cbar{{\overline C}}
\def\C{\mathbf{C}}
\def \red#1{\textcolor{red}{#1}}
\def \blue#1{\textcolor{blue}{#1}}
\hyphenation{nik-ol-oul-op-oul-os}
\newcommand{\widesim}[2][1.5]{
  \mathrel{\overset{#2}{\scalebox{#1}[1]{$\sim$}}}
}

\title{COUPLING COUPLES WITH COPULAS: ANALYSIS OF ASSORTATIVE MATCHING ON RISK ATTITUDE}
\date{}
\author{
Aristidis K. Nikoloulopoulos\footnote{{\small\texttt{A.Nikoloulopoulos@uea.ac.uk}}, School of Computing Sciences, University of East Anglia, Norwich NR4 7TJ, UK} \and  Peter G. Moffatt\footnote{{\small\texttt{P.Moffatt@uea.ac.uk}}, School of Economics, University of East Anglia, Norwich NR4 7TJ, UK}}
\maketitle

\begin{abstract}
\baselineskip=23pt
\noindent 
We investigate patterns of assortative matching on risk attitude, using self-reported (ordinal) data on risk attitudes for males and females within married couples, from the German Socio-Economic Panel over the period 2004-2012.  We apply a novel copula-based bivariate panel ordinal model.  Estimation is in two steps: firstly, a copula-based Markov model is used to relate the marginal distribution of the response in different time periods, separately for males and females; secondly, another copula is used to couple the males' and females' conditional (on the past) distributions. We find positive dependence, both in the middle of the distribution, and in the joint tails, and we interpret this as positive assortative matching (PAM).  Hence we reject standard assortative matching theories based on risk-sharing assumptions, and favour models based on alternative assumptions such as the ability of agents to control income risk. We also find evidence of ``assimilation"; that is, PAM appearing to increase with years of marriage.
\\


\end{abstract}

\section{Introduction}

\baselineskip=23pt

In recent years, the Economic Theory literature has seen intense interest in the concept of assortative matching on risk attitude, particularly when applied to the marriage market.  One obvious reason for this interest is that the nature of this type of matching has profound implications for the relationship between individual decision making and household decision making (see \citealt{browning_chiappori_weiss_2014}).  Much of this literature is based on risk-sharing arguments, which amount to the substitutability of risk-bearing between individuals in an incomplete insurance market: a risk-averse female is a demanding buyer of insurance, while a risk-seeking male is a ready seller of it \citep{Schulhofer-Wohl-2006, Legros&Newman2007, Chiappori-Reny-2006}.  This approach leads to the unambiguous prediction of negative assortative matching (NAM): the most risk-averse male will match with the least risk-averse female; the second most risk-averse male will match with the second least risk-averse female; and so on.

Against this background, a number of other theorists have demonstrated that the introduction of certain other model features can reverse the prediction to one of positive assortative matching (PAM).  For example, \cite{Li-etal2013} propose a model in which agents can control the risks to their incomes (by e.g. re-training, changing career or taking a second job).  They first show that if agents can only control the mean of the income distribution, the matching formula contains only a risk-sharing effect which is negative, and NAM is predicted.  However, they then show that if agents can control both the mean and the variance, there is in addition a risk-management effect which is positive, and if this term outweighs the risk-sharing effect, PAM is predicted.  The intuition underlying this prediction is that agents prefer similar partners because of their aligned objectives in risk management.  \cite{Gierlinger&Laczo2017} find that assortative matching behaviour depends on the level of commitment, with PAM arising in a situation of limited commitment.  The assumption of limited commitment must be seen as highly plausible in the context of the marriage market, in which formal risk-sharing contracts are typically not signed.  \cite{Li-etal-2016} show that when risks are large compared with individuals' risk-free incomes, PAM may result.

Empirical and experimental evidence that is currently available is broadly favourable to PAM.  \cite{Bacon-etal-2014}  used repeated data on married couples within the German Socio-Economic Panel to investigate spousal correlation in risk attitude.  They applied the bivariate panel ordered probit model to the self-reported risk attitude data from this source.  They found that the individual-specific effects (in the risk attitude equation) for the two members of a married couple, are positively correlated, and this was interpreted as evidence of PAM.  Evidence of PAM has also been found in factors that are known to determine risk attitude, for example education, wages and wealth \citep{becker1974,lam1988,Charles&Hurst2003}, and also in factors relating to non-financial risk-taking such as smoking \citep{Clark2006}.  \cite{DiCagno-etal-2012} carry out experiments in which agents allocate their wealth among risky lotteries and share winnings with their partners according to pre-committed rules.  Again, evidence of PAM is found: agents tend to choose partners with similar risk attitude to themselves.

Clearly this body of empirical evidence has important implications for the validity of each of the various assortative matching theories cited above.  Given this, it is imperative that the econometric approaches used to establish such evidence are valid, and sufficiently flexible to detect whatever  patterns of assortative matching exist within the data.  With this in mind, in this paper we delve deeper into the investigation of assortative mating on risk attitude, by applying the copula approach. The main advantage of the copula approach is that it can allow a wide range of flexible tail dependence and asymmetry between the two variables (here male's and female's risk attitude) under investigation.

It is this aspect of the copula approach that makes it particularly well-suited to the empirical testing of some of the most recent assortative matching theories.  This is because the theories not only predict PAM, but go further than this, by predicting that PAM is stronger in particular parts of the distribution of risk attitudes.  For example, in the version of the model of \cite{Li-etal2013} in which agents can control both the mean and the variance of their income distribution, highly risk averse agents are more strongly motivated to match with similar agents, hence implying positive tail dependence in the joint distribution of risk aversion.  This is a consequence of risk aversion of the ``household representative agent" becoming more sensitive to the choice of a dissimilar partner when risk aversion increases.  Even more recently, \cite{Chen-etal-2018} have explored similar issues in the context of a principal-agent problem with moral hazard: the agent can exert effort in risk reduction, but this effort is unobservable to the principal.  They find that a highly risk-averse principal is willing to pay more than a less risk-averse principal to avoid matching with a less risk-averse agent.  In other words, PAM is expected to be stronger at higher levels of risk aversion, once again implying positive tail dependence in the joint distribution of risk aversion.  The impact of moral hazard in matching problems is also considered by \cite{Wang-2013}.

Note that the copula approach is ideal for the empirical testing of the theories outlined in the preceding paragraph, since it explicitly embodies the feature of tail dependence  \citep{joe93} that is predicted by each of these theories.  Note also that existing empirical approaches used to investigate assortative matching, such as the random effects models used by  \cite{Clark2006} and \cite{Bacon-etal-2014}, are based on the normality assumption, and although mathematically convenient, are constrained to tail independence.  A further attraction of the copula approach is that the richness of the dependence structure is being introduced without the requirement of estimation of additional parameters.

We are also interested in whether and how this dependence structure changes with years of marriage.  A very interesting hypothesis is that of assimilation: a process by which attitudes become more similar with years of marriage. \cite{Bacon-etal-2014} found evidence of spousal correlation in risk attitude increasing with years of marriage, and \cite{DiFalco&Vieider2017}, using a sample of Ethiopian married couples, find that assimilation is  more important than assortative matching in explaining risk attitudes.  The hypothesis of assimilation will be  investigated in this paper by estimating the copula model separately for different stages of marriage. 

We will use a general copula construction, based on a set of bivariate copulas to jointly model bivariate ordinal time-series responses with covariates. We call this the ``joint copula-based Markov model''. For the proposed model, we construct the joint distribution in terms of three bivariate copulas. For each ordinal time series we consider a copula-based Markov model, where a parametric copula family is used for the joint distribution of subsequent observations and then we relate these ordinal time-series responses using another copula to couple their conditional (on the past) distributions at each time point.  Much is known about properties of parametric bivariate copula families in terms of dependence and tail behavior. In this paper we draw on this wealth of knowledge. 
The bivariate, or multinomial probit model,  in the biostatistics \citep{Ashford&Sowden1970}, psychometrics \citep{Muthen1978}, and econometrics \citep{Hausman&Wise1978} literature, is a simple example of the bivariate normal (BVN) copula with univariate probit regressions as the marginals.
Other choices of copulas are better if (a)  $Y_j$'s have more probability in
joint upper
or lower tail than would be expected with a discretized BVN, or (b)
$Y_j$'s can be considered as discretized maxima/minima or mixtures of discretized means
rather than discretized means \citep{nikoloulopoulos&joe12}.  Copulas that arise from extreme value theory have more probability in one joint tail (upper
or lower) than expected with a discretized BVN distribution or an BVN copula with discrete margins. It is also possible that there can be more probability in both the joint upper and joint lower tail, compared with discretized BVN models. This happens if the respondents consist of a ``mixture" population (e.g., different ethnicities).  From the theory of elliptical distributions and copulas, it is known that some scale mixtures of BVN have more dependence in the tails.

Much is known about copulas. They have been used to great benefit in many areas (e.g., insurance, finance and hydrology), but have only recently taken hold in econometric research.  So far in the Economics literature, the copula approach has been used to model earnings, mobility, house prices, and measures of well-being \citep{Dardanoni&Lamber2001,Bonhomme&Robin2006,
Bonhomme&Robin2009,Zimmer2012,zimmer15,Decancq2014}.  To our knowledge, our paper is the first application of the copula approach to an assortative matching problem.

The remainder of the paper proceeds as follows. Section \ref{overview} has a brief overview of relevant copula theory.  Section \ref{themodel} introduces
the joint copula-based Markov model for discrete ordinal responses with covariates and discusses its relationship with existing models.
 Section \ref{sec-families} discusses  suitable  parametric families of copulas for the joint copula-based Markov model for discrete ordinal responses with covariates. Estimation techniques and computational details are provided in Section \ref{estimation}.  
Section \ref{sec-appl} presents the applications
of our methodology to assortative mating using data from the German Socio-Economic Panel over the period 2004--2012.
We conclude with some discussion in Section \ref{sec-discussion}.

\section{\label{overview}Overview and relevant background for copulas}
A copula is a multivariate cumulative distribution function (cdf) with uniform $U(0,1)$ margins \citep{joe97,nelsen06,joe2014}.
If $F$ is a bivariate cdf with univariate margins $F_1,F_2$,
then Sklar's (1959) \nocite{sklar1959} theorem implies that there is a copula $C$ such that
\begin{equation}\label{copulacdf}
F(y_1,y_2)= C\Bigl(F_1(y_1),F_2(y_2)\Bigr).
\end{equation}
The copula is unique if $F_1,F_2$ are continuous, but not
if some of the $F_j$ have discrete components.
If $F$ is continuous and $(Y_1,Y_2)\sim F$, then the unique copula
is the distribution of $(U_1,U_2)=\left(F_1(Y_1),F_2(Y_2)\right)$ leading to
  $$C(u_1,u_2)=F\Bigl(F_1^{-1}(u_1),F_2^{-1}(u_2)\Bigr),
  \quad 0\le u_j\le 1, j=1,2,$$
where $F_j^{-1}$ are inverse cdfs. In particular, if $\Phi_{2}(\cdot;\rho)$
is the BVN cdf with correlation $\rho$ and
standard normal margins, and $\Phi$ is the univariate standard normal cdf,
then the BVN copula is
$$
C(u_1,u_2)=\Phi_{2}\Bigl(\Phi^{-1}(u_1),\Phi^{-1}(u_2);\rho\Bigr).
$$
The major advantage of copulas for dependence modelling is that the dependence structure may be separated from the univariate margins; see, for example, Section 1.6 of \cite{joe97}.
If $C(\cdot;\theta)$ is a parametric
family of copulas and $F_j(\cdot;\eta_j)$ is a parametric model for the
$j$th univariate margin, then
  $$C\Bigl(F_1(y_1;\eta_1),F_2(y_2;\eta_2);\theta\Bigr)$$
is a bivariate parametric model with univariate margins $F_1,F_2$.
For copula models, the variables can be continuous or discrete      \citep{Nikoloulopoulos2013a,nikoloulopoulos&joe12}.

\section{\label{themodel}A joint copula-based Markov model}
For ease of exposition, let $T$ be the dimension of a 
``panel" and $n$ the number of clusters. The theory can
be extended to varying cluster sizes.
Let $p$ be the number of
covariates, that is, the dimension of a covariate vector $\x$.
Let $Y^\star\sim \mathcal{F}$ be a latent variable,  such that $Y=y$ if
$\alpha_{y-1}+\x^T\bbf\leq Y^\star\leq  \alpha_{y}+\x^T\bbf,\,y=1,\ldots,K,$
where $K$ is the number of categories of $Y$
(without loss of generality, assume $\alpha_0=-\infty$ and  $\alpha_K=\infty$), and $\bbf$ is the $p$-dimensional regression vector.
From this definition, the response $Y$ is assumed to have density
$$f(y;\mu,\gbf)=\mathcal{F}(\alpha_{y}+\mu)-\mathcal{F}(\alpha_{y-1}+\mu),$$
where $\mu=\x^T\bbf$ is a function of $\x$
and the $p$-dimensional regression vector $\bbf$, and $\gbf=(\alpha_1,\ldots,\alpha_{K-1})$ is the $q$-dimensional vector of the univariate cutpoints ($q=K-1$). Note that $\mathcal{F}$ normal leads to the ordinal probit model for ordinal response, $\mathcal{F}$ logistic
leads to the ordinal cumulative logit model.

Suppose that data are $(y_{itj}, \x_{itj}),\, i = 1, . . . ,n,\,t=1,\ldots,T,\, j=1,2$
where $i$ is an index for individuals or clusters, $t$ is an index for
the repeated measurements or within cluster measurements, and $j$ is an index for the gender: 1=male, 2=female.
The univariate marginal model for
$Y_{itj} $ is $f_j(y_{itj}; \mu_{itj},\gbf_j)$ where  $\mu_{itj}=\x_{itj}^\top\bbf_j$ and  $\gbf_j$ of dimension $q_j$ be the vector of univariate cutpoints. If for each $t$, $Y_{i1j},\ldots,Y_{iTj}$ are serially independent conditional on $\mu_{itj}$, then the
log-likelihood for each gender is
\begin{eqnarray}\label{indlik}\ell_{j}= \sum_{i=1}^n\sum_{t=1}^T\, \log f_j(y_{itj};\mu_{itj},\gbf_j).
\end{eqnarray}

If the ordinal data are observed in a time series sequence as above, then the ordinal regression model can be adapted in two ways:
\begin{itemize}
\item add lagged variables as covariates; 
\item make use of time series models for stationary ordinal data. 
\end{itemize} 
Here, we use the second of these.  
For dependent $Y_{i1j},\ldots,Y_{iTj}$, estimation of $\bbf_j$ and $\gbf_j$ involves copula-based Markov models \citep[page 244]{joe97} for ordinal time-series with covariates. The joint distribution of adjacent observations is modelled through a parametric copula family from which transition probabilities may be extracted.
 The advantages of using time-series models are \citep{Joe-2015-proceedings}:

\begin{itemize}
\itemsep=0pt

\item The class of autocorrelation functions is much wider than those based on an ordered probit with lagged dependent variables appearing as explanatory variables.

\item Prediction in regressions with time dependent observations is simpler as they can be formulated with or without the preceding observations.

\item Serial dependence (positive or negative) can be modelled through suitable copula families.

\item 
The conditional expectation is generally non-linear  and a variety patterns are possible.  For extreme values of the variables, conditional expectation and variance are determined by the tail behaviour of the copula family.

\item Incorporating covariates in time-series models is more straightforward in univariate regression models.

\item 
It is straightforward to extend first-order Markov to higher orders.

\item 
Likelihood inference is easy provided the copula family has relatively simple form.

\end{itemize}
 Note in passing that \cite{chen-fan-06} studied copula-based Markov models of continuous response data. 

The transition cdf of $Y_{tj}$ given $Y_{t-1,j}$ is
\begin{eqnarray*}F_{j|t}(y_{tj}|y_{t-1,j})&=&P(Y_{tj}\leq
y_{tj}|Y_{t-1,j}=y_{t-1,j})\\&=&\Bigl[C_{j|t}\bigl(F(y_{t-1,j}),F(y_{tj})\bigr)-C_{j|t}\bigl(F(y_{t-1,j}-1),F(y_{tj})\bigr)\Bigr]/f(y_{t-1,j}),
\end{eqnarray*}
and the transition probability mass function (pmf) is
$$f_{j|t}(y_{tj}|y_{t-1,j})=P(Y_{tj}=y_{tj}|Y_{t-1,j}=y_{t-1,j})=\frac{f(y_{tj},y_{t-1,j})}{f_j(y_{t-1,j})},$$
where
$f(y_t,y_{t-1})=C_{j|t}\bigl(F(y_t),F(y_{t-1})\bigr)-C_{j|t}\bigl(F(y_t-1),F(y_{t-1})\bigr)-C_{j|t}\bigl(F(y_t),F(y_{t-1}-1)\bigr)+C_{j|t}\bigl(F(y_t-1),F(y_{t-1}-1)\bigr)$.
Then the log-likelihood for each gender is 
\begin{equation}\label{serlik}\ell_{j|t}= \sum_{i=1}^n\left(\log f_j(y_{i1j};\mu_{i1j},\gbf_j)+\sum_{t=2}^T\, \log f_{j|t}(y_{itj}|y_{i,t-1,j};\mu_{itj},\mu_{i,t-1,j},\gbf_j)\right).
\end{equation}

The BVN copula is a special case, and this is called ``autoregressive-to-anything" in \cite{biller&nelson2005} as acknowledged by \cite{joe2014}. Other copulas would be useful for the transition probability if there is more clustering of consecutive large or small values than would be expected with BVN.

Note in passing  that even though parameter estimates from univariate analysis ignoring the serial dependence remain consistent but they are inefficient  \citep{liang&zeger86}. When using copula terms in the likelihood improves asymptotic efficiency over the independence estimating equations \citep{Prokhorov&Schmidt2009}.

So far we treat the ordinal response of the males and the ordinal response of the females separately as if they were independent. In the sequel, we novel propose to relate these responses using a copula to couple their conditional (on the past) distributions at each time point. One should perform the analysis not ignoring serial dependence. The valid copula  practice is to first fit univariate time series models to correct for serial dependence as above and in the sequel join the  conditional joint distributions of each time-series using a multivariate copula; see e.g. \cite{Patton-2012}.

From \cite{sklar1959}, there is a
bivariate copula $C_{12|t}$
such that $\Pr(Y_{t1}\le y_{t1}, Y_{t2}\le y_{t2})=C_{12|t}\bigl(F_{1|t}(y_{t1}|y_{t-1,1}),F_{2|t}(y_{t2}|y_{t-1,2})\bigr)$. Then it follows that the joint pmf is
\begin{align}\label{jointserpmf}
&f_{12|t}(y_{t1},y_{t2})=\\&C_{12|t}\bigl(F_{1|t}(y_{t1}|y_{t-1,1}),F_{2|t}(y_{t2}|y_{t-1,2})\bigr)-C_{12|t}\bigl(F_{1|t}(y_{t1}-1|y_{t-1,1}),F_{2|t}(y_{t2}|y_{t-1,2})\bigr)-\nonumber\\&C_{12|t}\bigl(F_{1|t}(y_{t1}|y_{t-1,1}),F_{2|t}(y_{t2}-1|y_{t-1,2})\bigr)+C_{12|t}\bigl(F_{1|t}(y_{t1}-1|y_{t-1,1}),F_{2|t}(y_{t2}-1|y_{t-1,2})\bigr)\nonumber
\end{align}

For the joint copula-based Markov model, we let $C_{1|t},C_{2|t}$ and
$C_{12|t}$ be parametric bivariate copulas, say with parameters $\theta_1,\theta_2$
and $\theta$, respectively.
For the set of all parameters, let $\thbf=\{\bbf_j,\gbf_j,\theta_j,\theta: j=1,2\}$.
We model
the joint distribution in terms of three  bivariate
copulas.
There is much known about properties of parametric bivariate copula families
in terms of dependence and tail behavior.
Note that the copula $C_{j|t}$ models the time-series for the  $j$th response and the copula $C_{12|t}$ links the ordinal
response for males to the ordinal response for females. 
Our general statistical model allows for selection of $C_{j|t}$ and $C_{12|t}$ independently among a variety of parametric copula families, i.e.,
there are no constraints in the choices of parametric copulas $\{C_{j|t},C_{12|t}:
j=1,2\}$.

\section{\label{sec-families}Choices of parametric families of copulas}

\begin{figure}[!h]
\begin{center}
\begin{tabular}{cc}
\includegraphics[width=0.3\textwidth]{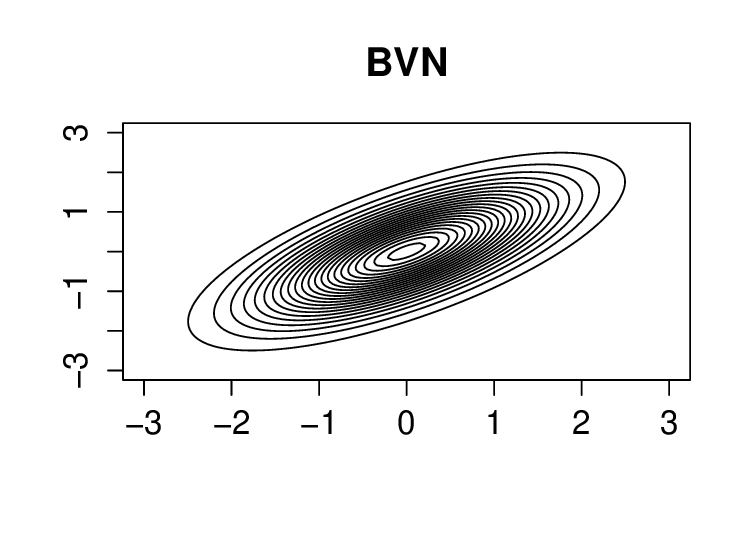}
&

\includegraphics[width=0.3\textwidth]{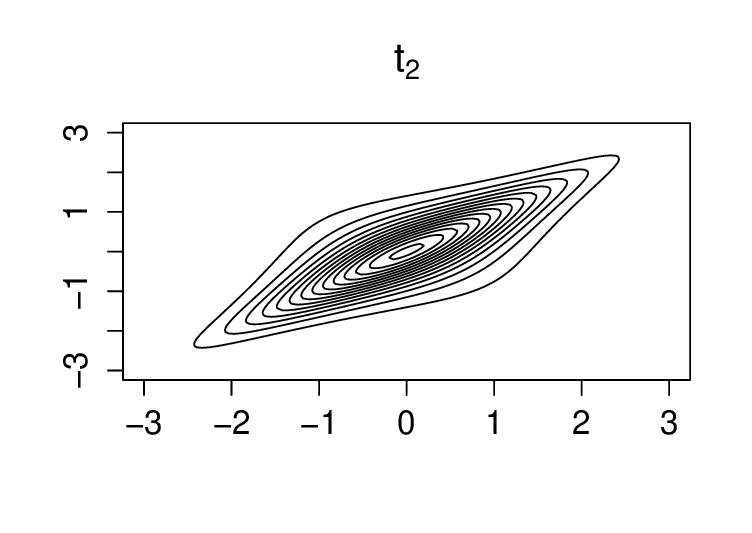}\\

\multicolumn{2}{c}{\includegraphics[width=0.3\textwidth]{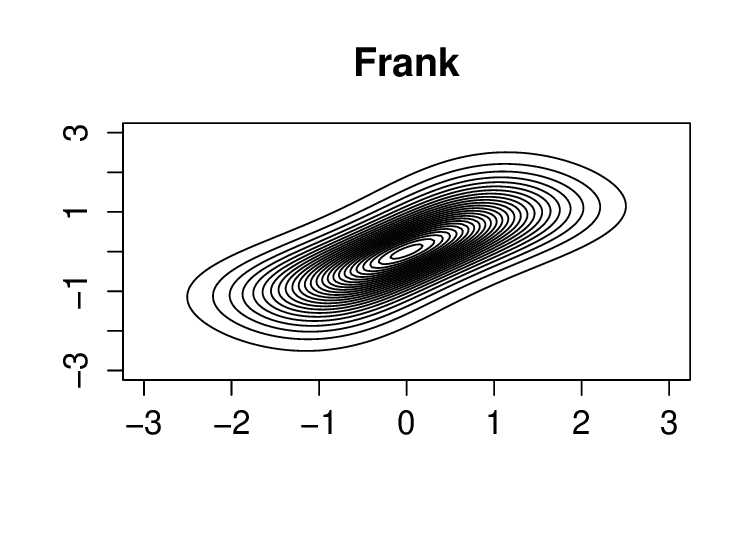}}
\\
\includegraphics[width=0.3\textwidth]{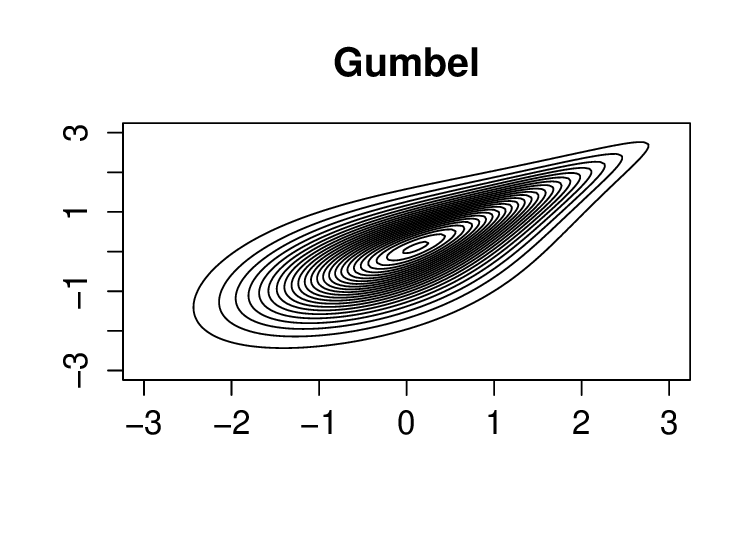}
&
\includegraphics[width=0.3\textwidth]{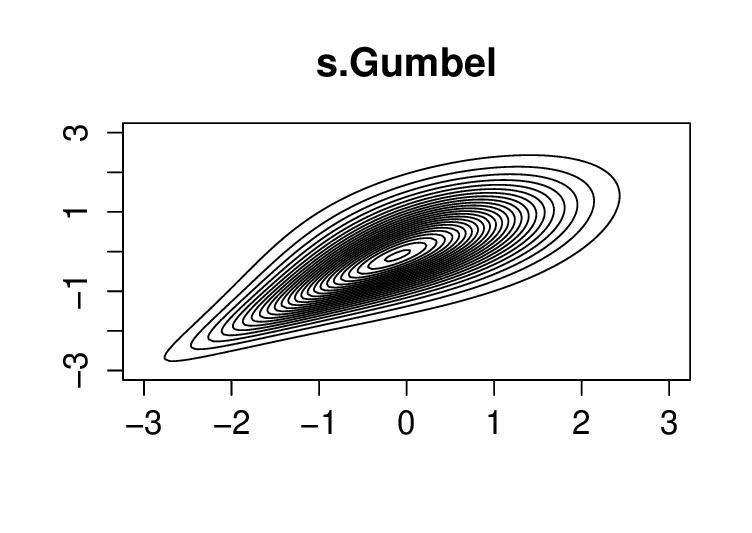}
\\
\end{tabular}
\caption{\label{contours}Contour plots of BVN, $t_2$, Frank, Gumbel and s.Gumbel copulas with standard normal margins and dependence   parameters  corresponding to Kendall's $\tau$ value of $0.5$. }
\end{center}
\end{figure}

In our candidate set, families that have
different strengths of tail behaviour (see e.g., \cite{nikoloulopoulos&joe&li11,nikoloulopoulos&joe12}) are included. In the descriptions below, a bivariate copula $C$ is {\it reflection symmetric}
if its density 
 satisfies $c(u_1,u_2)=c(1-u_1,1-u_2)$ for all $0\leq u_1,u_2\leq 1$.
Otherwise, it is reflection asymmetric often with more probability in the
joint upper tail or joint lower tail. {\it Upper tail dependence} means
that $c(1-u,1-u)=O(u^{-1})$ as $u\to 0$ and {\it lower tail dependence}
means that $c(u,u)=O(u^{-1})$ as $u\to 0$.
If $(U_1,U_2)\sim C$ for a bivariate copula $C$, then $(1-U_1,1-U_2)\sim
C_{180^0}$, where $C_{180^0}(u_1,u_2)=u_1+u_2-1+C(1-u_1,1-u_2)$ is the survival (or rotated by 180 degrees) copula of $C$; this ``reflection"
of each uniform $U(0,1)$ random variable about $1/2$ changes the direction
of tail asymmetry.
\begin{itemize}
\item
Reflection symmetric copulas with tail independence satisfying
$C(u,u)=O(u^2)$ and $\Cbar(1-u,1-u)=O(u^2)$ as $u\to 0$,
such as the Frank copula with cdf
$$C(u_1,u_2;\th)=-\theta^{-1}\log \left\{1+\frac{(e^{-\theta u_1}-1)(e^{-\theta
u_2}-1)}{e^{-\theta}-1} \right\},\quad \theta \in (-\infty,\infty)\setminus\{0\}.$$
\item Reflection asymmetric copula family with upper tail dependence,
such as the Gumbel extreme value copula with cdf
$$C(u_1,u_2;\th)=\exp\Bigl[-\Bigl\{(-\log u_1)^{\theta}
+(-\log u_2)^{\theta}\Bigr\}^{1/\theta}\Bigr],\quad \th\geq 1.$$
The resulting model in this case  has  more probability in the joint upper tail compared to the BVN copula. That is  there is more dependence of  large ordinal values that would be expected with BVN.

\item Reflection asymmetric copula family with  lower tail dependence, such as the
survival Gumbel (s.Gumbel) copula with cdf
$$C(u_1,u_2;\th)=u_1+u_2-1 + \exp\Bigl[-\Bigl\{\bigl(-\log (1-u_1)\bigr)^{\theta}
+\bigl(-\log (1-u_2)\bigr)^{\theta}\Bigr\}^{1/\theta}\Bigr],\quad \th\geq 1.$$
The resulting model in this case  has more probability in the joint lower tail compared to the BVN copula. That is  there is more dependence of  small ordinal values that would be expected with BVN. 
\item\label{t} Copulas with reflection symmetric upper and lower tail
dependence, such as the bivariate Student $t_\nu$ copula with cdf
$$C(u_1,u_2;\th)=T_2\Bigl(T^{-1}(u_1;\nu),T^{-1}(u_2;\nu);\th,\nu\Bigr),\quad-1\leq\th\leq 1,$$
where $T(;\nu)$ is the univariate Student t cdf with (non-integer) $\nu$ degrees of freedom, and $T_2$ is the
cdf of a bivariate Student t distribution with $\nu$ degrees of freedom and correlation parameter $\th$.
A small value of $\nu$, such as $1\le \nu\le 5$, leads to a model with
more probabilities in the joint upper and joint lower tails compared to the BVN copula. That is  there is more dependence of  large and small ordinal values than would be expected with BVN. 
\end{itemize}
To depict the concept of reflection symmetric (asymmetric) tail dependence (independence), we plot contour plots of the corresponding copula densities with standard normal margins and dependence   parameters  corresponding to Kendall's $\tau$ value of $0.5$ in Figure \ref{contours}. 

For this paper, the above copula families are sufficient for the applications in Section \ref{sec-appl}, since tail dependence is a property to consider when choosing amongst different families of copulas and the concept of upper/lower tail dependence is one way to differentiate families.  \cite{Nikoloulopoulos&karlis08CSDA}
have shown that it is hard to choose a copula with similar properties from real data, since
copulas with similar (tail) dependence properties provide similar fit. 
Note also that these copulas satisfy the conditions under which a copula function generates a stationary Markov chain that satisfies mixing conditions at a geometric rate \citep{chen-etal-2009,Beare2010}.

\section{\label{estimation}Estimation techniques and computational details}
The log-likelihood of the joint copula-based Markov model is
\begin{equation}\label{jointserlik}\ell_{12|t}(\thbf)= \sum_{i=1}^n\left(\log f_{12}(y_{i11},y_{i12};\thbf) + \sum_{t=2}^T\, \log f_{12|t}(y_{it1},y_{it2};\thbf)\right).
\end{equation}
where $f_{12|t}(\cdot)$ is given in (\ref{jointserpmf}) and 
\begin{align*}
&f_{12}(y_{11},y_{12})=\\&C_{12|t}\bigl(F_{1}(y_{11}),F_{2}(y_{12})\bigr)-C_{12|t}\bigl(F_{1}(y_{11}-1),F_{2}(y_{12})\bigr)-\nonumber\\&C_{12|t}\bigl(F_{1}(y_{11}),F_{2}(y_{12}-1)\bigr)+C_{12|t}\bigl(F_{1}(y_{11}-1),F_{2}(y_{12}-1)\bigr).
\end{align*}

The maximum likelihood (ML) estimates can be derived using the steps below:

\begin{enumerate}
\item For each $j$: 
\begin{enumerate}
\item At the first step the $\ell_j(\bbf_j,\gbf_j)$ in (\ref{indlik}) is maximized over the univariate marginal parameters $\bbf_j,\gbf_j$ assuming time independence.

\item At the second step the $\ell_{j|t}(\bbf_j,\gbf_j,\th_j)$ in (\ref{serlik}) is maximized over 
the copula parameter $\th_j$
with univariate  parameters $\bbf_j,\gbf_j$ fixed as estimated at  the first step.
\item At the third step the $\ell_{j|t}(\bbf_j,\gbf_j,\th_j)$ in (\ref{serlik}) is maximized over both the univariate parameters $\bbf_j,\gbf_j$ and 
copula parameter $\th_j$
with initial parameters the estimates for the preceding  steps.

\end{enumerate}  
\item At the fourth step the $\ell_{12|t}(\thbf)$ in (\ref{jointserlik}) is maximized over 
the copula parameter $\th$
with all the other parameters fixed as estimated at  the preceding steps.

\item At the final step the $\ell_{12|t}(\thbf)$ in (\ref{jointserlik}) is maximized over $\thbf$
with initial parameters the estimates for the preceding  steps. 
\end{enumerate}
In the steps above   
the Inference function of
Margins method  \citep{joe97,joe05} is used to get initial estimates. 

Each of the  estimated parameters can be obtained by 
using a quasi-Newton \citep{nash90} method applied to the log-likelihood.  
This numerical  method requires only the objective
function, i.e., the joint log-likelihood, while the gradients
are computed numerically and the Hessian matrix of the second
order derivatives is updated in each iteration. The standard errors (SEs) of the ML estimates can be also obtained via the gradients and the Hessian computed numerically during the maximization process. Assuming that the usual regularity conditions \citep{serfling80} for
asymptotic maximum likelihood theory hold for the bivariate model
as well as for its margins we have that ML estimates are
asymptotically normal. Therefore one can build Wald tests to
statistically judge the effect of any covariate.

\section{\label{sec-appl}Application to the German Socio-Economic Panel}

Couples were followed during 2004--2012.  Interviews were administered 
every 2 years between 2004 and 2008, and every year thereafter, resulting in  7 measurements per couple.  We use the 2201 couples observed at all seven time points, although our methodology could be applied to situations in which the cluster size $T$ is non-constant.

The outcome variable in our analysis is the risk attitude question which is asked in each of the seven survey years.  The question is reproduced as follows:

\begin{quote}
How do you see yourself?  Are you generally a person who is fully prepared to take risks or do you try to avoid taking risks?  Please tick a box on the scale, 
where 0 means ``risk averse" and 10 means ``fully prepared to take risks"
\end{quote}

\begin{center}
\begin{tabular}{ccccccccccc}
$\Box$&$\Box$&$\Box$&$\Box$&$\Box$&$\Box$&$\Box$&$\Box$&$\Box$&$\Box$&$\Box$\\
0&1&2&3&4&5&6&7&8&9&10\\
\end{tabular}
\end{center}

\bigskip

\noindent We also use a number of explanatory variables: number of children, log of household income, age, age-squared and education level.

We fit the joint copula-based Markov  model with BVN,
Gumbel, s.Gumbel, and $t_\nu$ bivariate linking copulas.
For Student $t_\nu$, choices of $\nu$ were $1,2,\ldots,10$. For the  model we allow three different copula
families, one for the male time-series, one for the female time-series and one to join them.
To make it easier
to compare the dependence parameters, we convert the estimated parameters to Kendall's $\tau$'s
in $(0,1)$  via the relations 
$\tau=\frac{2}{\pi}\arcsin{\theta}$,
$\tau=1+4\theta^{-1}\left[\frac{1}{\theta}\int_0^{\theta}\frac{t}{e^t-1}dt-1\right],$
and 
$\tau=1-\th^{-1}$
for elliptical, Frank and Gumbel copulas
in \cite{HultLindskog02}, \cite{genest87}, and \cite{genest&mackay86}, respectively.
Note that Kendall's {$\tau$} only accounts for the dependence dominated by the middle of the data, and it is expected to be similar amongst  different families of copulas. However, the tail dependence varies, as explained in Section \ref{sec-families}, and is a property to consider when choosing amongst different families of copulas.
For the  model with $t_\nu$ we summarize the
choice of integer $\nu$ with the largest maximized log-likelihood.

{Since the number of parameters is the same between the models, we  use the  log-likelihood at  estimates as a  measure for goodness of fit between all the  models.
We further compute the \nocite{vuong1989}Vuong's  (1989) test to  check if there is
more probability in the joint tails than the one expected via  assuming a BVN copula to couple the conditional (on the past) distributions of male and female ordinal time-series.    
The Vuong's test is the sample version of the difference in Kullback-Leibler divergence between two models and can be used to differentiate two  parametric models which could be non-nested. 
Assume that we have Models 1 and 2 with parametric densities $f^{(1)}_{12|t}$ and  $f^{(2)}_{12|t}$  with   $C_{12|t}$ being the BVN copula and any other parametric family of copulas with different tail properties, respectively; the best fit of $C_{j|t}$ is used for the joint distribution of subsequent observations  for males and females. 
The sample version  of the difference in Kullback-Leibler divergence between two models with MLEs $\hat\thbf^{(1)},\hat\thbf^{(2)}$ is
$$\bar D=\sum_{i=1}^N D_i/N,$$
where $D_i=\log\left[\frac{f^{(2)}_{12|t}\left(Y_1,Y_2;\hat\thbf^{(2)}\right)}{f^{(1)}_{12|t}\left(Y_1,Y_2;\hat\thbf^{(1)}\right)}\right]$.
Model 1  is the better fitting model if $\bar D<0$, and Model 2 is the better fitting model if $\bar D>0$. 
\cite{vuong1989} 
has shown that asymptotically under the null hypothesis $H_0:\Delta=0$, i.e.,  Models 1 and 2 have the same  parametric densities $f^{(1)}$ and  $f^{(2)}$,  
$$z_0=\sqrt{N}\bar D/s\widesim{H_0}\mathcal{N}(0,1),$$
where  $s^2=\frac{1}{N-1}\sum_{i=1}^N(D_i-\bar D)^2$. 
 For more details we refer the interested  reader to  
\cite{joe2014,Nikoloulopoulos2015b}.  
}

For these risk data, if a respondent reports the maximum (minimum) willingness to take risk in year $t$, then it seems natural to expect them to report the maximum (minimum) in year $t-1$ and year $t+1$ as well.
That is, based on the data descriptions, 
we can expect {\it a priori} that a  model with $C_{j|t}$ being the $t_\nu$ copulas might
be  plausible, as in this case the data have more probability in the joint tails. 
Furthermore, since the sample is a mixture (males and females)  we can expect {\it a priori} that a  $t_\nu$ copula to join  the male and female time-series might be plausible, as in this case the ordinal responses can be considered as mixtures of discretized means.

Table \ref{time-series} contains results from the copula-based Markov models  for ordinal time series with covariates for both males and females, where a parametric copula family $C_{j|t}$ is used for the joint distribution of subsequent observations estimated separately for males and females, using data from all available years.  On the evidence of the maximised log-likelihoods, it is clear that the $t_\nu$ copula with a small $\nu$ is the best-fitting model for these data, and, there is a big improvement over the  ``autoregressive-to-anything" (BVN copula-based Markov) model.  In particular, we find that the $t_4$ copula is the best-fitting model for the male time series, while the $t_5$ is the best-fitting for female time series.

\begin{sidewaystable}[htbp]
  \centering 
  \caption{Estimated parameters, and joint log-likelihoods
$\ell_{j|t}$ for the copula-based Markov models  for ordinal time series with covariates for both males and females, where a parametric copula family $C_{j|t}$ is used for the joint distribution of subsequent observations for 2004--12 time period.}
\bigskip    
    \begin{tabular}{ccccccccccc}
    \toprule
   
   $C_{j|t}$       & \multicolumn{2}{c}{BVN} & \multicolumn{2}{c}{Frank} & \multicolumn{2}{c}{Gumbel} & \multicolumn{2}{c}{s.Gumbel} & $t_4$ & $t_5$ \\
          & Males  & Females & Males  & Females & Males  & Females & Males  & Females & Males  & Females \\
           
          \hline

    $\a_1$ & -0.708 & -1.381 & -0.803 & -1.257 & -1.259 & -1.839 & -0.105 & -1.143 & -0.595 & -1.507 \\
    $\a_2$ & -0.244 & -0.861 & -0.296 & -0.719 & -0.754 & -1.293 & 0.269 & -0.704 & -0.165 & -1.010 \\
    $\a_3$ & 0.352 & -0.249 & 0.317 & -0.104 & -0.130 & -0.663 & 0.781 & -0.147 & 0.401 & -0.408 \\
    $\a_4$ & 0.818 & 0.258 & 0.780 & 0.390 & 0.349 & -0.150 & 1.210 & 0.341 & 0.859 & 0.099 \\
    $\a_5$ & 1.126 & 0.600 & 1.077 & 0.721 & 0.661 & 0.191 & 1.505 & 0.681 & 1.166 & 0.444 \\
    $\a_6$ & 1.719 & 1.256 & 1.647 & 1.363 & 1.249 & 0.825 & 2.094 & 1.346 & 1.762 & 1.103 \\
    $\a_7$ & 2.116 & 1.664 & 2.034 & 1.773 & 1.630 & 1.201 & 2.497 & 1.765 & 2.160 & 1.508 \\
    $\a_8$ & 2.703 & 2.193 & 2.622 & 2.314 & 2.172 & 1.666 & 3.102 & 2.316 & 2.743 & 2.029 \\
    $\a_9$ & 3.478 & 2.867 & 3.402 & 3.013 & 2.845 & 2.224 & 3.907 & 3.028 & 3.494 & 2.680 \\
    $\a_{10}$ & 3.972 & 3.263 & 3.905 & 3.422 & 3.252 & 2.540 & 4.431 & 3.449 & 3.964 & 3.055 \\
    \# children & -0.013 & -0.027 & -0.011 & -0.020 & -0.002 & -0.021 & -0.007 & -0.027 & 0.001 & -0.024 \\
    $\log$(HH income) & 0.174 & 0.005 & 0.166 & 0.009 & 0.137 & -0.009 & 0.192 & 0.005 & 0.172 & 0.001 \\
    Age   & -0.109 & 0.090 & -0.104 & 0.124 & -0.112 & 0.004 & -0.074 & 0.106 & -0.086 & 0.044 \\
    Age$^2$ & 0.003 & -0.012 & 0.004 & -0.014 & 0.004 & -0.004 & 0.000 & -0.013 & 0.001 & -0.008 \\
    Education & 0.008 & 0.028 & 0.008 & 0.027 & 0.000 & 0.025 & 0.017 & 0.033 & 0.007 & 0.030 \\
    $\tau_j$ & 0.308 & 0.289 & 0.347 & 0.322 & 0.336 & 0.314 & 0.334 & 0.307 & 0.334 & 0.309 \\\hline
    $\ell_{j|t}$ & -30734.0 & -30434.6 & -30605.8 & -30335.6 & -30552.8 & -30332.9 & -30694.0 & -30407.5 & -30445.4 & -30225.1 \\
 
    \bottomrule
    \end{tabular}%
  \label{time-series}%

\begin{footnotesize}
\begin{flushleft}
Age and Age$^2$ are scaled by factor $10^{-1}$ and  $10^{-2}$ respectively. 
\end{flushleft}
\end{footnotesize}
\end{sidewaystable}%

On this basis, these two copulas are chosen for the joint copula-based Markov model that couples the two univariate ordinal time series.  For the joint copula-based Markov model  for ordinal time series with covariates for both males and females, where a $t_4$ and a $t_5$ copula family is used for the joint distribution of subsequent observations for males and females, once again, a number of different copulas are tried   to form  the joint distribution of couple observations, and the results from these joint models are presented in Table \ref{joint-copula}.  This time, we find that the $t_5$ copula provides the best fit. {In fact,  from the Vuong's statistic there is enough improvement  {compared to the BVN copula} to get a highly  statistical significant difference ($p$-value$<0.001$).  
This result suggests some skewness to both upper and lower tail for the pair of male and female risks.}

\begin{sidewaystable}[htbp]

  \centering
  \caption{Estimated parameters, their standard errors, joint log-likelihoods 
$\ell_{12|t}$, and Vuong's statistics for the joint copula-based Markov models  for ordinal time series with covariates for both males and females, where a $t_4$ and a $t_5$ copula family is used for the joint distribution of subsequent observations for males and females, respectively, and, an additional parametric copula family $C_{12|t}$ is used for the joint distribution of couple observations for the 2004--12 time period.}
\bigskip    
    \begin{tabular}{ccccccccccc}
    \toprule
   
  $C_{12|t}$        & \multicolumn{2}{c}{BVN} & \multicolumn{2}{c}{Frank} & \multicolumn{2}{c}{Gumbel} & \multicolumn{2}{c}{s.Gumbel} & \multicolumn{2}{c}{$t_5$} \\
          & Males  & Females & Males  & Females & Males  & Females & Males  & Females & Males  & Females \\
           
          \hline
    {$\a_1$} & {-0.611} & {-1.508} & {-0.601} & {-1.508} & {-0.595} & {-1.494} & {-0.687} & {-1.625} & -0.612 (0.282)     & -1.507 (0.276) \\
    {$\a_2$} & {-0.171} & {-1.005} & {-0.168} & {-1.008} & {-0.158} & {-0.990} & {-0.245} & {-1.124} & -0.164 (0.282)     & -0.999 (0.276) \\
    {$\a_3$} & {0.405} & {-0.399} & {0.404} & {-0.403} & {0.416} & {-0.383} & {0.327} & {-0.521} & 0.414 (0.282)     & -0.390 (0.276) \\
    {$\a_4$} & {0.868} & {0.107} & {0.867} & {0.103} & {0.879} & {0.125} & {0.786} & {-0.016} & 0.875 (0.283)     & 0.118 (0.276) \\
    {$\a_5$} & {1.176} & {0.451} & {1.175} & {0.445} & {1.188} & {0.468} & {1.093} & {0.328} & 1.183 (0.283)     & 0.460 (0.276) \\
    {$\a_6$} & {1.772} & {1.105} & {1.770} & {1.098} & {1.781} & {1.117} & {1.688} & {0.985} & 1.775 (0.283)     & 1.111 (0.276) \\
    {$\a_7$} & {2.169} & {1.506} & {2.165} & {1.501} & {2.174} & {1.511} & {2.086} & {1.391} & 2.169 (0.284)     & 1.509 (0.276) \\
    {$\a_8$} & {2.747} & {2.023} & {2.745} & {2.019} & {2.742} & {2.014} & {2.669} & {1.913} & 2.742 (0.284)     & 2.018 (0.277) \\
    {$\a_9$} & {3.488} & {2.669} & {3.494} & {2.669} & {3.461} & {2.638} & {3.420} & {2.567} & 3.475 (0.285)     & 2.652 (0.278) \\
    {$\a_{10}$} & {3.952} & {3.041} & {3.965} & {3.045} & {3.902} & {2.997} & {3.892} & {2.945} & 3.930 (0.287)     & 3.019 (0.280) \\
    {\# children} & {0.002} & {-0.022} & {0.005} & {-0.021} & {0.001} & {-0.028} & {0.002} & {-0.023} & 0.003 (0.015)     & -0.028 (0.015) \\
    {$\log$(HH income)} & {0.179} & {0.005} & {0.174} & {0.003} & {0.175} & {0.010} & {0.172} & {0.003} & 0.174 (0.022)     & 0.012 (0.021) \\
    {Age} & {-0.109} & {0.033} & {-0.099} & {0.037} & {-0.088} & {0.024} & {-0.114} & {-0.003} & -0.086 (0.074)     & 0.020 (0.074) \\
    {Age$^2$} & {0.003} & {-0.007} & {0.002} & {-0.007} & {0.001} & {-0.007} & {0.004} & {-0.003} & 0.002 (0.006)     & -0.006 (0.007) \\
    {Education} & {0.006} & {0.029} & {0.007} & {0.029} & {0.007} & {0.030} & {0.005} & {0.027} & 0.005 (0.004)     & 0.027 (0.004) \\
    {$\tau_j$} & {0.330} & {0.311} & {0.327} & {0.309} & {0.334} & {0.313} & {0.329} & {0.310} & 0.333 (0.006)     & 0.312 (0.006) \\
    {$\tau$} & \multicolumn{2}{c}{0.171} & \multicolumn{2}{c}{0.176} & \multicolumn{2}{c}{0.160} & \multicolumn{2}{c}{0.173} & \multicolumn{2}{c}{0.172 (0.006)}   \\\hline
    {$\ell_{12|t}$} & \multicolumn{2}{c}{-60145.5} & \multicolumn{2}{c}{-60197.1} & \multicolumn{2}{c}{-60173.9} & \multicolumn{2}{c}{-59987.1} & \multicolumn{2}{c}{-59942.2} \\\hline
 
Vuong's & $z_0$&$p$-value& $z_0$&$p$-value& $z_0$&$p$-value& $z_0$&$p$-value&$z_0$&$p$-value \\
test &\multicolumn{2}{c}{-}& -3.792    &$<0.001$ & -1.817  & 0.069& 8.536&$<0.001$ &7.831 &$<0.001$\\
    \bottomrule
    \end{tabular}%
  \label{joint-copula}%
\begin{footnotesize}
\begin{flushleft}
Standard errors  are shown in parentheses for the best fit; Age and Age$^2$ are scaled by factor $10^{-1}$ and  $10^{-2}$ respectively. 
\end{flushleft}
\end{footnotesize}
\end{sidewaystable}%

In the final (double) column of Table \ref{joint-copula}, we provide estimates together with SEs for this joint model.  The coefficients associated with the explanatory variables lead us to the following conclusions: for males, income has a strong positive effect on willingness to take risk; for females, number of children has  a negative effect, while education has a positive effect. Of course, it may be that some of these explanatory variables are endogenous.  It is well known that risk attitude is a key determinant in the child-bearing decision \citep{Schmidt2008}, in the education decision \citep{Shaw96}, and also consequently in the determination of income.  The significance of the coefficients on these variables in our model must surely be partly explained by this reverse causality.  However, the fact that child-bearing and education appear important only in the female equation, while income appears important only in the male equation is interesting and calls for further investigation.

The estimate of Kendall's $\tau$ appearing in the final column of Table \ref{joint-copula} is 0.172 and this is strongly significant. In fact, a joint copula-based Markov model  leads to better inferences than a copula-based Markov model with independence of males and females since the likelihood has been improved by $728.3=-59942.2-(-30445.4-30225.1)$. This indicates that there is strong evidence of positive dependence between males and females in the middle of the distribution (i.e. at normal levels of risk attitude).  The fact that the best-fitting copula for the joint model is $t_5$ (instead of say, BVN) indicates that there is also positive tail dependence: risk-lovers are particularly keen to match with other risk-lovers, and risk-avoiders with risk-avoiders. This is confirmed by the Vuong's  statistic of 7.831 (p-value$<0.001$) reported in the final row of Table \ref{joint-copula}, which establishes clear superiority of the $t_5$ over the BVN.

Next, we have carried out estimation separately for two different time ranges: 2004--8 and 2009--12.  Table \ref{separate} shows results from the univariate models using data from 2004--8 and   2009--12 separately, from which we conclude that the $t_\nu$ copula  with a small $\nu$ provides the best-fitting model for both univariate ordinal time-series in both time periods.

Table \ref{separate} also shows results from the joint models estimated with data from 2004--8 and 2009--12 separately, and assuming the univariate models found to be the best-fitting as described above.  Comparing the results for the two time periods, we note two key differences.  Firstly, the estimate of Kendall's $\tau$ rises from 0.146 in 2004--8 to 0.191 in 2009--12, and moreover, it may be verified from the associated SEs that the two 95\% confidence intervals do not overlap.  Hence we have evidence that Kendall's $\tau$ rises with years of marriage.  The second key difference is that the best-fitting joint model for the 2004--8 data is $t_6$, while the best-fitting joint model for 2009--12 is $t_5$.  There is a reduction of one in the degrees of freedom of the best-fitting $t_\nu$ copula.  Both of these differences between the results 
provide evidence that dependence increases with years of marriage; both in the middle of the distribution, and in the joint tails.

\begin{sidewaystable}[htbp]
  \centering
  \caption{\label{separate} Estimated parameters 
and  log-likelihoods 
for the  best fitted copula-based Markov models and joint copula-based Markov models  for ordinal time series with covariates for both males and females for the 2004--8 and 2009--12 time periods. Standard errors are shown in the parentheses for the joint fit.  
}
    \begin{tabular}{ccccccccccccccc}
    \toprule
          &       & \multicolumn{6}{c}{2004--8}                    &       & \multicolumn{6}{c}{2009--12} \\
   \cmidrule{3-8} \cmidrule{11-15}
          &       &       & \multicolumn{2}{c}{Separate} &       & \multicolumn{2}{c}{Joint} &       &       & \multicolumn{2}{c}{Separate} &       & \multicolumn{2}{c}{Joint} \\
          &       &       & \multicolumn{2}{c}{$t_4$}       &       & \multicolumn{2}{c}{$t_6$} &       &       & $t_4$ & $t_5$ &       & \multicolumn{2}{c}{$t_5$} \\
           \cmidrule{3-5} \cmidrule{7-8} \cmidrule{11-12} \cmidrule{14-15}
          &       &       & Males  & Females &       & Males  & Females &       &       & Males  & Females &       & Males  & Females \\\midrule
        $\a_1$  &       &       & -0.474 & -1.312 &       &  -0.684 (0.369)      &  -1.381 (0.363)      &       &       & -0.046 & -0.771 &       &  -0.062 (0.423)      &  -0.930 (0.411)  \\
        $\a_2$  &       &       & 0.177 & -0.55 &       &  -0.024 (0.367)      &  -0.616 (0.361)      &       &       & 0.35  & -0.338 &       &  0.352 (0.423)      &  -0.483 (0.411)  \\
        $\a_3$  &       &       & 0.814 & 0.119 &       &  0.615 (0.367)      &  0.053 (0.362)      &       &       & 0.9   & 0.233 &       &  0.919 (0.424)      &  0.101 (0.411)  \\
        $\a_4$  &       &       & 1.257 & 0.642 &       &  1.059 (0.367)      &  0.575 (0.362)      &       &       & 1.373 & 0.728 &       &  1.400 (0.424)      &  0.601 (0.411)  \\
        $\a_5$  &       &       & 1.581 & 1.008 &       &  1.382 (0.368)      &  0.938 (0.362)      &       &       & 1.666 & 1.055 &       &  1.696 (0.424)      &  0.928 (0.412)  \\
        $\a_6$  &       &       & 2.165 & 1.655 &       &  1.963 (0.368)      &  1.580 (0.362)      &       &       & 2.262 & 1.718 &       &  2.292 (0.425)      &  1.583 (0.412)  \\
        $\a_7$  &       &       & 2.547 & 2.049 &       &  2.342 (0.369)      &  1.969 (0.363)      &       &       & 2.669 & 2.133 &       &  2.693 (0.426)      &  1.989 (0.412)  \\
        $\a_8$  &       &       & 3.102 & 2.565 &       &  2.892 (0.370)      &  2.479 (0.364)      &       &       & 3.277 & 2.658 &       &  3.287 (0.426)      &  2.500 (0.413)  \\
        $\a_9$  &       &       & 3.862 & 3.208 &       &  3.640 (0.372)      &  3.113 (0.366)      &       &       & 4.027 & 3.313 &       &  4.016 (0.428)      &  3.136 (0.415)  \\
        $\a_{10}$  &       &       & 4.362 & 3.647 &       &  4.129 (0.374)      &  3.536 (0.370)      &       &       & 4.478 & 3.604 &       &  4.456 (0.431)      &  3.425 (0.418)  \\
        \# children  &       &       & 0.006 & -0.039 &       &  0.005 (0.019)      &  -0.039 (0.019)      &       &       & 0.006 & 0.01  &       &  0.011 (0.022)      &  -0.002 (0.022)  \\
        $\log$(HH income)  &       &       & 0.207 & 0.051 &       &  0.203 (0.031)      &  0.061 (0.031)      &       &       & 0.18  & -0.004 &       &  0.177 (0.029)      &  0.004 (0.028)  \\
        Age    &       &       & -0.071 & 0.059 &       &  -0.118 (0.100)      &  0.014 (0.097)      &       &       & 0.008 & 0.219 &       &  0.020 (0.113)      &  0.159 (0.114)  \\
        Age$^2$  &       &       & 0.000     & -0.009 &       &  0.005 (0.009)      &  -0.005 (0.009)      &       &       & -0.004 & -0.018 &       &  -0.005 (0.009)      &  -0.014 (0.010)  \\
        Education  &       &       & 0.008 & 0.034 &       &  0.004 (0.006)      &  0.030 (0.006)      &       &       & 0.004 & 0.023 &       &  0.005 (0.005)      &  0.022 (0.006)  \\
        $\tau_j$  &       &       & 0.308 & 0.266 &       &  0.314 (0.010)      &  0.279 (0.010)      &       &       & 0.365 & 0.355 &       &  0.360 (0.008)      &  0.352 (0.008)  \\
        $\tau$  &       &       & \multicolumn{2}{c}{ -     } &       & \multicolumn{2}{c}{ 0.146 (0.009)     } &       &       & \multicolumn{2}{c}{ -     } &       & \multicolumn{2}{c}{ 0.191 (0.008)     } \\\midrule
    Log-likelihood  &       &       & -13244.7 & -13062.8 &       & \multicolumn{2}{c}{-26078.4} &       &       & -17356.6 & -17198.8 &       & \multicolumn{2}{c}{-34074.5} \\
    \bottomrule
    \end{tabular}%
  \label{tab:addlabel}%
\end{sidewaystable}

At this point it is useful to recall that the sample used in this study is restricted to couples observed in every year.  Hence attrition (poorly-matched couples dropping out of the sample) is not an issue.  This means that the increase in dependence seen in the comparison of Tables 4 and 6 may be interpreted as clear evidence of ``assimilation'': the two members of the couple become more similar in terms of risk-attitude as their marriage progresses. This finding is in agreement with those of \cite{DiFalco&Vieider2017}.

Of course, this effect could simply be a time effect. For example, the global financial crisis occurred between our two sample ranges. Clearly this sort of  event has the potential  to influence  risk attitudes. However, it is less clear that this sort of event, and more generally the passage of time, has the potential to change  {\it the dependence patterns}  in risk attitude between spouses. We are therefore led to believe that our assimilation explanation is the most plausible.

\section{\label{sec-discussion}Discussion}

Assortative matching is a concept of great interest to economists, as evidenced by the extensive theoretical and empirical literatures devoted to it, dating back to  \cite{becker1974}.  The econometric modelling of assortative matching is clearly a setting in which the precise nature of the dependence between two variables representing the relevant outcomes for the two members of the couple, becomes the central focus of the analysis. The copula approach provides the ideal framework for this sort of modelling, since it brings richness to the dependence structure. For example, it allows the modelling of both joint tails and moreover allows a variety of types of asymmetry in the dependence structure. 

In applying the copula approach to the problem of assortative matching on risk attitude, we have found evidence of  
PAM, in agreement with previous empirical work.  However, the copula approach has enabled us to find evidence of both centre dependence and tail dependence. That is, over the entire range of of risk attitudes, there is a tendency for individuals to match with other individuals with similar risk attitude, and this positive dependence is particularly marked in the tails.

The evidence of PAM amounts to a rejection of standard assortative matching theories based on risk-sharing assumptions, and the favouring models based on alternative assumptions such as the ability of agents to control income risk, and limited commitment in the marriage contract.  
Moreover, since we have found evidence of tail dependence, our results may be seen to be consistent with the predictions of Li et al. (2013) and Chen et al. (2018), both of whom, as mentioned in Section 1, predict tail dependence (at least in the risk-averse tail).

These conclusions have been arrived at using a novel model which we have labelled the ``joint copula-based Markov model". The model consists of two stages. In the first stage, males and females are considered separately, and a copula is used to model the joint distribution of neighbouring (in time) outcomes for a given individual. The best fitting copulas thus found are then combined using a third copula that couples the two conditional distributions at each time point. In all three cases, the best-fitting  copula is found to be a $t_\nu$  with a small $\nu$, which leads to a model with more probability in the joint upper and joint lower tails compared to the BVN copula. 
This fact leads to the conclusion of positive dependence in the tails, and we have found strong statistical evidence to confirm  this, in the form of the Vuong's statistic.

Having arrived at this conclusion, we then extended the analysis by applying the same two-stage procedure separately for the 2004-8 and 2009-12 data. The key results here were that both middle and tail dependence were stronger in the second time range. We interpreted this in terms of the phenomenon of ``assimilation": the two members of the couple become more similar in risk attitude with the accumulation of years of marriage.

Our key conclusion is that the copula approach is very well-suited to the empirical investigation of assortative matching problems, and we look forward to discovering what sorts of dependence structures are found when matching on something other than risk attitude.

\section*{Acknowledgement}
We acknowledge access to the German Socio-Economic Panel for use in this research under licence number: 2596. Thanks to Phil Bacon for outstanding data management.


\end{document}